\documentclass{article}

\usepackage{hyperref}

\usepackage{multirow}

\usepackage{graphicx}

\usepackage{amsmath}
\usepackage{amssymb}
\usepackage{amsthm}

\newcommand{\bfr}{\mathbf{r}}
\newcommand{\bfu}{\mathbf{u}}

\newcommand*{\Ab}{\overline{A}}
\newcommand*{\Abb}{\overline{\overline{A}}}

\begin{document}

\title{Assignment of fields from particles to mesh}

\author{Daniel Duque \& Pep Espa\~nol \\
  Model Basin Research Group (CEHINAV). \\
  ETSI Navales, Universidad Polit\'ecnica de Madrid \\
  Madrid, Spain \\
  \href{mailto:daniel.duque@upm.es}{daniel.duque@upm.es} \\
  \& \\
  Dpto. de F{\'\i}sica Fundamental \\
  Universidad Nacional de Educaci\'on a Distancia \\
  Madrid, Spain
}

\date{\today}

\maketitle

\begin{abstract}
  In Computational Fluid Dynamics there have been many attempts to
  combine the power of a fixed mesh on which to carry out spatial
  calculations with that of a set of particles that moves following
  the velocity field. These ideas indeed go back to Particle-in-Cell
  methods, proposed about 60 years ago.
  Of course, some procedure is needed to transfer field information
  between particles and mesh. There are many possible choices for this
  ``assignment'', or ``projection''. 
  Several requirements may guide this choice. Two well-known ones are
  conservativity and stability, which apply to volume integrals of the
  fields.
  An additional one is here considered: preservation of
  information. This means that mesh interpolation, followed by mesh
  assignment, should leave the field values invariant. The resulting
  methods are termed ``mass'' assignments due to their strong
  similarities with the Finite Element Method.
  We test several procedures, including the well-known FLIP, on three
  scenarios: simple 1D convection, 2D convection of Zalesak's disk,
  and a CFD simulation of the Taylor-Green periodic vortex sheet. The
  most symmetric mass assignment is seen to be clearly superior to
  other methods.
\end{abstract}

\section{Introduction}
\label{sec:intro}

Historically, there have been two points of view to consider the
dynamics of fluids: Eulerian, describing the dynamics of fluid with
respect to an external frame, and Lagrangian, describing its dynamics
within the flow. These approached would be reflected in computational
fluid dynamics (CFD), in which the discretization may be carried out
on a fixed, Eulerian, mesh, or on a Lagrangian moving set of
computational particles. Each approach has its advantages and
drawbacks.

The idea of combining both approaches goes back to Particle-in-Cell
(PIC) methods, introduced in the 1950s \cite{PIC, evans1957,
  PIC2}. The hope was to carry out the numerically expensive
calculations related to spatial derivatives on a fixed mesh. The
velocity field would then be transferred to particles, which would
move according to it, advecting other fields. These would be
transferred back to the mesh to begin the next iteration.

This very appealing idea also has its drawbacks. The obvious one is
that a new procedure is needed to transfer the field information
between particles and mesh. This procedure has been termed
``assignment''.  We will here use ``projection'' as a synonym. Our
main task is to study the performance of different projection
techniques, under the guidance of several requirements. One of them is
conservativity: the total volume integral of a field must not vary
upon projection. Another is stability: the integral of the square of
any field should decrease upon projection. These two are well known,
while here we consider an additional one: preservation of
information. This means that the field values at discrete points do
not vary under projection interpolation followed by projection.

The article is organized as follows. In Section \ref{sec:assignment}
assignment procedures are discussed. First, conservativity and
stability are defined in Section \ref{sec:cons_stab}. Then, the
simplest assignment is seen introduced in
\ref{sec:delta_assignment}. The FLIP procedure, which is currently
very popular in the computer graphics community \cite{bridson_2015} is
considered in \ref{sec:flip_assignment}.  In Section
\ref{sec:mass_assignment} we provide a discussion of the mass
assignment idea, and its possible variants.
These procedures are then tested in three scenarios in Section
\ref{sec:results}. The first one, in Section \ref{sec:1D}, is a very
simple 1D convection of a top-hat function. This is actually a 1D
version of the well known Zalesak's disk 2D test, which is considered
in \ref{sec:zalesak}. Finally, a CFD simulation of the Taylor-Green
periodic vortex sheet, a solution of the Navier-Stokes equations, is
given in Section \ref{sec:TG}.
Some finishing remarks are given in Section \ref{sec:conclusions}.

\section{Assignment procedures}
\label{sec:assignment}


The set of functions $\{\psi_\mu\}$ is used to interpolate, or
reconstruct, a field $A(\bfr)$ from its values at particles $A_\mu$:
\begin{equation}
\label{eq:part_interp}
A(\bfr) \doteq \sum_\mu A_\mu \psi_\mu ( \bfr ) .
\end{equation}

The functions are supposed to comply with partition of unity, in order
a particle distribution with constant values yields a constant field:
\begin{equation}
\label{eq:sum_psi}
\sum_\mu \psi_\mu(\bfr) = 1 .
\end{equation}
In this work, we will only consider the simple linear Finite Element
basis functions (FEs), even though other choices are of course
possible.

For the mesh we will keep the same symbols, but the subindices will be
Latin letters. A field is reconstructed from the values at mesh nodes
$\Ab_i$ by mesh functions $\{\psi_i\}$:
\begin{equation}
\label{eq:mesh_interp}
\Ab(\bfr) \doteq \sum_i \Ab_i \psi_i ( \bfr ) .
\end{equation}

The particle-to-mesh assignment procedure consists of finding nodal
values for $\Ab_i$ given particle values $A_\mu$. The inverse
mesh-to-particle yields particle $\Abb_\mu$ given mesh $\Ab_i$.


\subsection{Conservativity and stability}
\label{sec:cons_stab}

An property that may lead us on our research of possible assignment
methods is conservativity. This expresses that the integral of a field
will not change upon projection. Integrating field $A$ of
Eq. (\ref{eq:part_interp}),
\[
\int A(\bfr) d\bfr  =  \sum_\mu A_\mu v_\mu ,
\]
where particle volumes are given by
\begin{equation}
  \label{eq:part_v}
  v_\mu := \int \psi_\mu(\bfr) d\bfr .
\end{equation}

Equivalently, mesh volumes may be defined from
Eq. (\ref{eq:mesh_interp}),
\begin{equation}
  \label{eq:mesh_v}
  v_i := \int \psi_i(\bfr) d\bfr .
\end{equation}
However, the FLIP method below does \emph{not} use this mesh volume.

Whichever the particular definition of the volumes, conservativity is
expressed as:
\begin{equation}
\label{eq:cons}
\sum_i \Ab_i v_i = \sum_\mu A_\mu v_\mu .
\end{equation}

In the case of linear momentum components, this would guarantee the
conservation of linear momentum. For a density field, this would
guarantee conservation of mass. It therefore would seem as a vital
requirement. However, we will see that methods that do not comply with
this condition still may deviate little from it.

Of course, the same requirement could be asked when projecting from the
mesh onto the particles:
\begin{equation}
\label{eq:cons2}
\sum_\mu \Abb_\mu v_\mu  = \sum_i \Ab_i v_i .
\end{equation}

Another requirement is stability: for any energy-like expression
defined on the particles and on the mesh:
\[
E_\mathrm{p} := \sum_\mu v_\mu A_\mu^2 \qquad
E_\mathrm{m} := \sum_i v_i \Ab_i^2 ,
\]
we require
\begin{equation}
\label{eq:stab}
E_\mathrm{m} \le E_\mathrm{p} .
\end{equation}
This guarantees that there is no overshoot in e.g. the kinetic energy
upon assignment. Also, for a general field this forces a diminishing
second momentum, which prevents overshooting in a global sense. Of
course, the same could be required when going from the mesh to the
particles.

\subsection{$\delta$ assignment}
\label{sec:delta_assignment}

The simplest assignment would be to define mesh values as the local
values reconstructed from the particles:
\begin{equation}
\label{eq:delta}
\Ab_i = \Ab(\bfr_i) = \sum_\mu A_\mu \psi_\mu( \bfr_i ) .
\end{equation}
Also,
\begin{equation}
\label{eq:delta2}
\Abb_\mu = \Abb(\bfr_i) = \sum_i \Ab_i \psi_i( \bfr_\mu ) .
\end{equation}

Particle volume may simply be defined as in (\ref{eq:part_v}) and
(\ref{eq:mesh_v}). It is easy to check that this procedure does not
satisfy either conservativity or stability.  In Table
\ref{table:methods} we will collect the relevant expressions for
the methods considered.

\begin{table}
  \centering
  \begin{tabular}{|c|c|c|}
    \hline
     method    &  part $\rightarrow$ mesh  & mesh $\rightarrow$ part    \\
     \hline
     \hline
     $\delta$  &
     $\displaystyle \Ab_i = \sum_\mu A_\mu \psi_\mu ( \bfr_i )$   (\ref{eq:delta} )  &
     \\
     FLIP &
     $\displaystyle \Ab_i :=  \sum_\mu A_\mu v_\mu \psi_i(\bfr_\mu) /\sum_\mu v_\mu \psi_i(\bfr_\mu)$ (\ref{eq:PIC},\ref{eq:FLIP_vol})
     &
     \multirow{3}{*}{$\displaystyle \Abb_\mu = \Abb(\bfr_i) = \sum_i \Ab_i \psi_i ( \bfr_\mu )$ (\ref{eq:delta2})}
     \\
     mass-$\delta$ &
     $\displaystyle \Ab_i := \sum_j m_{ij}^{-1} \int d\bfr A(\bfr) \phi_j(\bfr)$
     (\ref{eq:proj}, \ref{eq:phi_def}) &
     \\
     \hline
     full mass &
     $\displaystyle \Ab_i := \sum_j m_{ij}^{-1} \int d\bfr A(\bfr) \psi_j(\bfr)$
     (\ref{eq:proj}, \ref{eq:phi_def}) &
     $\displaystyle \Abb_\mu := \sum_\nu m_{\mu\nu}^{-1} \int d\bfr \Ab(\bfr) \psi_\nu(\bfr)$
     (\ref{eq:proj2}) \\
     \hline
     \hline
     mass - lumped &
     $\displaystyle \Ab_i := \sum_j m_{ij}^{-1} \int d\bfr A(\bfr) \psi_j(\bfr)$
     (\ref{eq:proj}, \ref{eq:phi_def}) &
     \multirow{2}{*}{$\displaystyle \Abb_\mu := \frac1{v_\mu} \int d\bfr \Ab(\bfr) \psi_\mu(\bfr)$}
     (\ref{eq:lumped}) \\
     lumped &
     $\displaystyle \Ab_i := \frac1{v_i} \int d\bfr A(\bfr) \psi_i(\bfr)$
     (\ref{eq:proj}, \ref{eq:lumped}) &
     \\
     \hline
  \end{tabular}
  \caption{Features of the methods considered (left column), the procedure
    by which fields are projected from particles to mesh (middle column), and the reverse
    procedure (right column). References are given to relevant equations in the text.
    The last two methods have been considered but results are not given.
    \label{table:methods}}
\end{table}

\subsection{FLIP assignment}
\label{sec:flip_assignment}

This procedure starts from the expression
\begin{equation}
\label{eq:PIC}
\Ab_i :=  \frac{1}{v_i} \sum_\mu A_\mu v_\mu \psi_i(\bfr_\mu) .
\end{equation}
The particle volumes may be defined as in (\ref{eq:part_v}).  If
(\ref{eq:mesh_v}) is used for the mesh volumes one, recovers the PIC
expression \cite{PIC, evans1957, PIC2}.

The FLIP procedure proposes the alternative mesh volume:
\begin{equation}
\label{eq:FLIP_vol}
v_i := \sum_\mu v_\mu \psi_i(\bfr_\mu) .
\end{equation}

This procedure can be shown to satisfy both conservativity and
stability.

A later projection onto the particles is exactly as in the
$\delta$-assignment, Eq. (\ref{eq:delta2}), and this can be seen to
again satisfy conservativity and stability.

This procedure may seem to be a great improvement since a simple
change in the mesh volume restores conservativity and stability.  It
is also convenient to code, since the mesh functional set,
$\{\psi_i\}$, is used for both particle to mesh assignment and its
reverse.  However, the volume of a nodal mesh (\ref{eq:FLIP_vol}) does
not depend on the mesh itself, but on the particles around it.  A node
may even have a vanishing volume if no particles are close and the
$\{\psi_\mu\}$ have compact support.  Also, the two assignments,
(\ref{eq:delta2}) and (\ref{eq:PIC}) are clearly unsymmetrical.
A summary of the method is given in Table \ref{table:methods}.

\subsection{Mass assignment}
\label{sec:mass_assignment}

Let us consider the assignment procedure
\begin{equation}
\label{eq:proj}
\Ab_i :=  \int d\bfr A(\bfr) \phi_i(\bfr) .
\end{equation}

Here, the assignment functional set $\{ \phi_i(\bfr) \}$ consists of
normalized functions:
\begin{equation}
\label{eq:delta_norm}
\int d\bfr \phi_i(\bfr) = 1 ,
\end{equation}
so that a constant field yields constant mesh values.  Notice the
$\{\psi_\mu\}$ functions carry no physical units, but the
$\{\phi_\mu\}$ have units of length$^{-d}$, where $d$ is the spatial
dimension.  The $\delta$ method is recovered as a special case if
Dirac $\delta$ functions are used,
$ \phi_i(\bfr) = \delta(\bfr-\bfr_i)$, which in retrospect explains
its name.

For conservativity, let us evaluate
\begin{equation}
\label{eq:cons_check}
\sum_i \Ab_i v_i =
\sum_i
\left(
  \int d\bfr A(\bfr) 
\right)
\phi_i(\bfr) v_i =
\int d\bfr A(\bfr)
\left(
 \sum_i \phi_i(\bfr) v_i 
\right) .
\end{equation}
If the last parenthesis was equal to $1$, conservativity would apply
(with particle volumes as in (\ref{eq:part_v}). Therefore, these
functions must satisfy
\begin{equation}
\label{eq:delta_sum}
\sum_i v_i \phi_i(\bfr) = 1 .
\end{equation}
If this condition holds, it is straightforward to proof that stability
is also satisfied.

The simplest way to define these functions would be as normalized
versions of the $\{ \psi_i \}$ :
\begin{equation}
\label{eq:lumped}
  \phi_i(\bfr) = \frac1{v_i}  \psi_i(\bfr) 
\end{equation}
This would correspond to a ``lumped mass'' method, a term that will
become clear very soon.

Let us however consider $\{\phi_i\}$ that ``preserves nodal
information''. By this, we mean that a reconstruction procedure,
followed by projection, should leave the nodal values invariant:
\begin{equation*}
\Ab_i :=  \int d\bfr  \phi_i(\bfr) \left(\sum_j \Ab_j \psi_j(\bfr) \right) .
\end{equation*}
(See also Ref. \cite{cottet2000} , p. 243 for an application of this
idea to an iterative method.)  Notice these two operations are carried
out on the mesh only (or, on particles only).  This means
\begin{equation}
\label{eq:ortho}
\int d\bfr \phi_i(\bfr) \psi_j(\bfr) = \delta_{ij} ,
\end{equation}
where the latter $\delta$ is Kronecker's. We will call this the
``preservation property''.  If the $\{\phi_i\}$ are linear
combinations of the $\{\psi_i\}$ set, it is simple to show that
\begin{equation}
\label{eq:phi_def}
\phi_i(\bfr) = \sum_j m_{ij}^{-1} \psi_j(\bfr) ,
\end{equation}
where the inverse of the mass matrix appears, the latter defined as
having elements
\[
m_{ij} =:  \int d\bfr  \psi_i(\bfr)   \psi_j(\bfr) .
\]

It can be shown that the functions defined by (\ref{eq:phi_def}) do
comply with requirements (\ref{eq:delta_norm}) and
(\ref{eq:delta_sum}). As a consequence, the resulting procedure will
be conservative and stable.

For the sake of symmetry, a similar procedure is employed for
projection onto the particles:
\begin{equation}
\label{eq:proj2}
\Abb_\mu :=  \int d\bfr \Abb(\bfr) \phi_\mu(\bfr) .
\end{equation}
The resulting procedure can also be shown to satisfy both
conservativity and stability.

Some remarks are in order.

First, the integration needed in (\ref{eq:proj}) may be cumbersome to
carry out. We therefore employ a simplified quadrature rule that
involves a quadratic interpolation for function $\Ab(\bfr)$, as
explained below in Appendix \ref{sec:quadrature}.

Second, this procedure requires matrix inversion. This is not such a
problem for the mesh, since in a typical CFD computation these
matrices must be assembled and inverted on the mesh anyway. On the
particles however, such a calculation is often not needed, whereas it
certainly is within the present procedure.

Third, the simple (\ref{eq:lumped}) would correspond to a lumped mass
approximation in the language of the Finite Element Method
(FEM). Indeed, $\sum_j m_{ij}= v_i$.

Fourth and last, there is an appealing correspondence with the usual
FEM. Indeed, beginning with a Poisson equation:
\[
g - \nabla^2 A  = 0 ,
\]
we may project onto a nodal functional space (it is immaterial for
this discussion whether the following occurs on the particles, or on
the mesh):
\[
\sum_i g_i \psi_i(\bfr)  -  \sum_i A_ i \nabla^2 \psi_i(\bfr)  \approx 0.
\]
The equality is no longer satisfied in general, but we may ask the
residual to be orthogonal to all the shape functions (as in the method
of weighted residuals \cite{reddy}):
\[
\int d\bfr
\left(
\sum_i g_i \psi_i(\bfr)  -  \sum_i A_ i \nabla^2 \psi_i(\bfr) 
\right)
\psi_j(\bfr) =0 \qquad \forall j .
\]
This results in the expression
\[
\sum_i m_{ij} g_i =
  \int d\bfr  \psi_j(\bfr)   \nabla^2 
\left(
  \sum_i A_i \psi_i(\bfr) 
\right) .
\]
Now, we may invert the mass matrix, and recalling our definition
(\ref{eq:phi_def}),
\[
 g_j =
 \int d\bfr  \phi_j(\bfr)   \nabla^2 
 \left(
   \sum_i A_i \psi_i(\bfr) 
 \right) = 
 \int d\bfr  \phi_j(\bfr)   \nabla^2 \Ab(\bfr)
\]
This is an expression for the second derivative that entails: the
reconstruction from the nodal values $A_i$ to a function $\Ab(\bfr)$,
deriving this function twice, then projecting back onto the nodes.  We
therefore see that a classical FEM approach yields an expression that
is consistent with our mass assignment. Indeed, if the method to solve
the equations of motion is of the FEM type, the calculations are
already likely implemented in the code (at least, for the mesh).

We will consider two instances of mass projection. In the first one,
mass projection will be used from the particles to the mesh, but
simple $\delta$ projection will be used when projecting back. This is
because in a typical pFEM-like simulation the matrices necessary for
the former projection will be likely already computed, at the start of
the simulation (they will not change since the mesh is fixed).  If its
performance was good, its numerical implementation would not imply a
large increase in computational resources. We will call this method
``mass-$\delta$'', for obvious reasons. The $\delta$ projection
is not conservative nor stable, but we will see below that
departs little from these requirements.

Our second choice will be ``full mass'' projection. This is the
most symmetric mass assignment method, which will comply with conservativity
and stability. It clearly requires the particle mass matrix must be
calculated, and inverted, at each time step. We will see that this additional
computational burden may be compensated by its superior performance.

Another possible candidate would be a
``mass-lumped'' method, where only the particle volumes are needed.
Of course, a purely ``lumped'' method, both-ways, can be
considered. These two lumped procedures are listed at the end of Table
\ref{table:methods}, but the results are not shown here, since in all
cases they are very similar to the mass-$\delta$ results. They
do satisfy conservativity and stability.

\section{Numerical experiments}
\label{sec:results}

\subsection{Moving top-hat in one dimension}
\label{sec:1D}

On the $[0,1)$ segment, let us consider a region with a colour field
$A$ that has a value of $1$ for $x \in (0.25,0.75)$ and $0$
otherwise. The field is simply advected:
\begin{equation}
  \frac{d A}{d t} = 0 \qquad   \frac{d x}{d t} = u .
\end{equation}
In our case, $u = 1$, simply a constant positive velocity (towards the right).

Since the field is just advected, it makes little sense to project
from and onto the mesh at every time-step. With no projection, the
shape translates to the right. Nevertheless, the $A$ field is
projected from the mesh to the particles and vice versa at every time
step, in order to benchmark our method. We employ $200$ particles and
$200$ mesh nodes, with a time step given by a Courant number
Co$:= u (\Delta t) / (\Delta x) = 0.1 $.

In the following, we will restrict our attention to four
procedures. The fist one the $\delta$ projection, which is the
simplest one. This was used in our previous study
\cite{duque_2016b}. We will also consider the assignment FLIP
method. To be precise, we are only evaluating the FLIP volume
assignment (\ref{eq:FLIP_vol}), the FLIP idea of projecting only
increments in fields at each time step does not apply here since there
is no source term to field $A$.

For the mesh, and also for the particles in the mass method, we use
simple linear FE functions. In the context of vortex methods, these
procedure is termed ``area-weighting scheme'', also Cloud-in-Cell
\cite{christiansen1973}. These are affine functions that are equal to
$1$ at the nodes, then linearly decrease to $0$ at the two neighbouring
nodes.

\begin{figure}
  \centering
  \begin{minipage}{0.45\textwidth}
      \includegraphics[width=\textwidth]{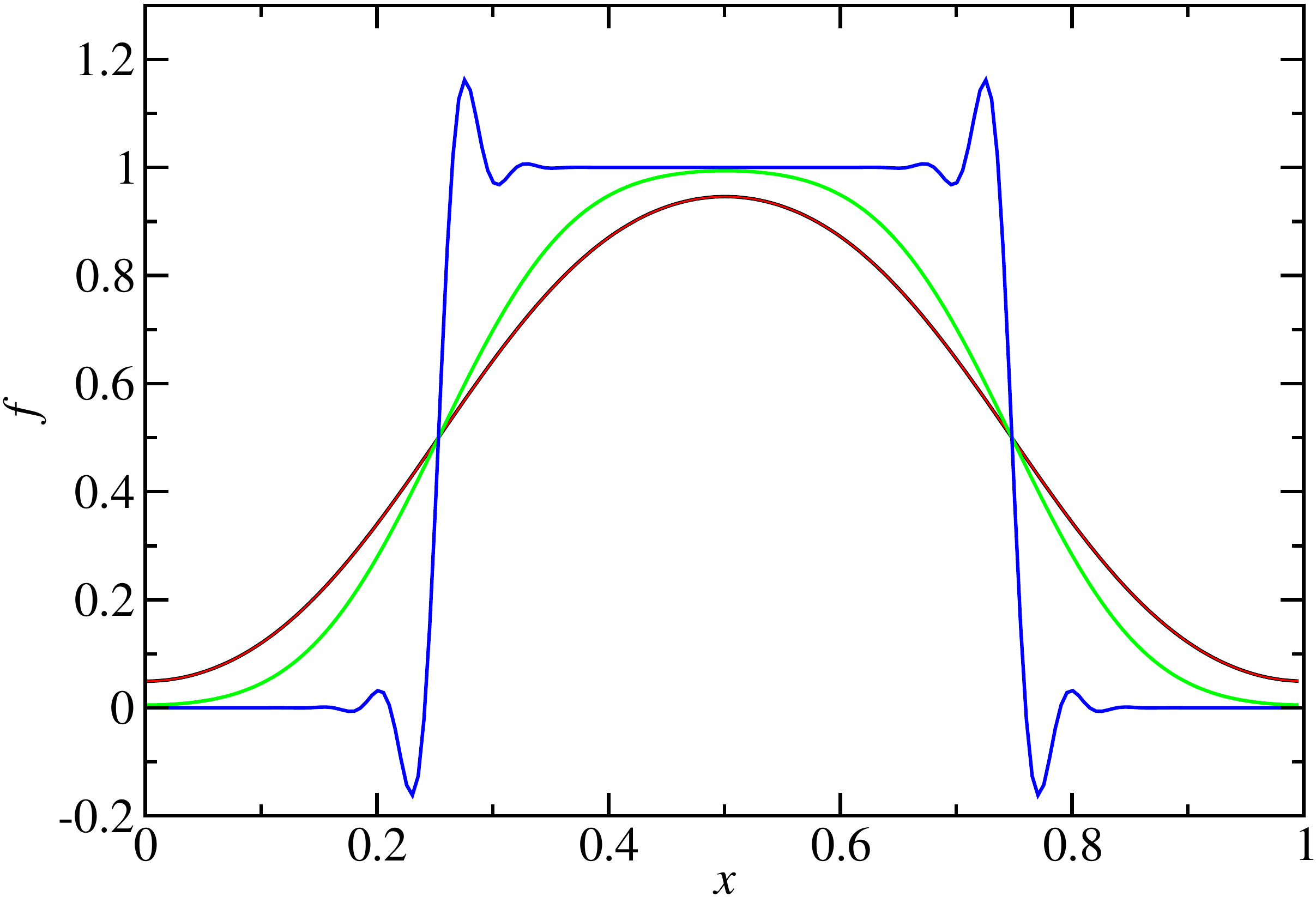}
  \end{minipage}
  \quad
  \begin{minipage}{0.45\textwidth}
      \includegraphics[width=\textwidth]{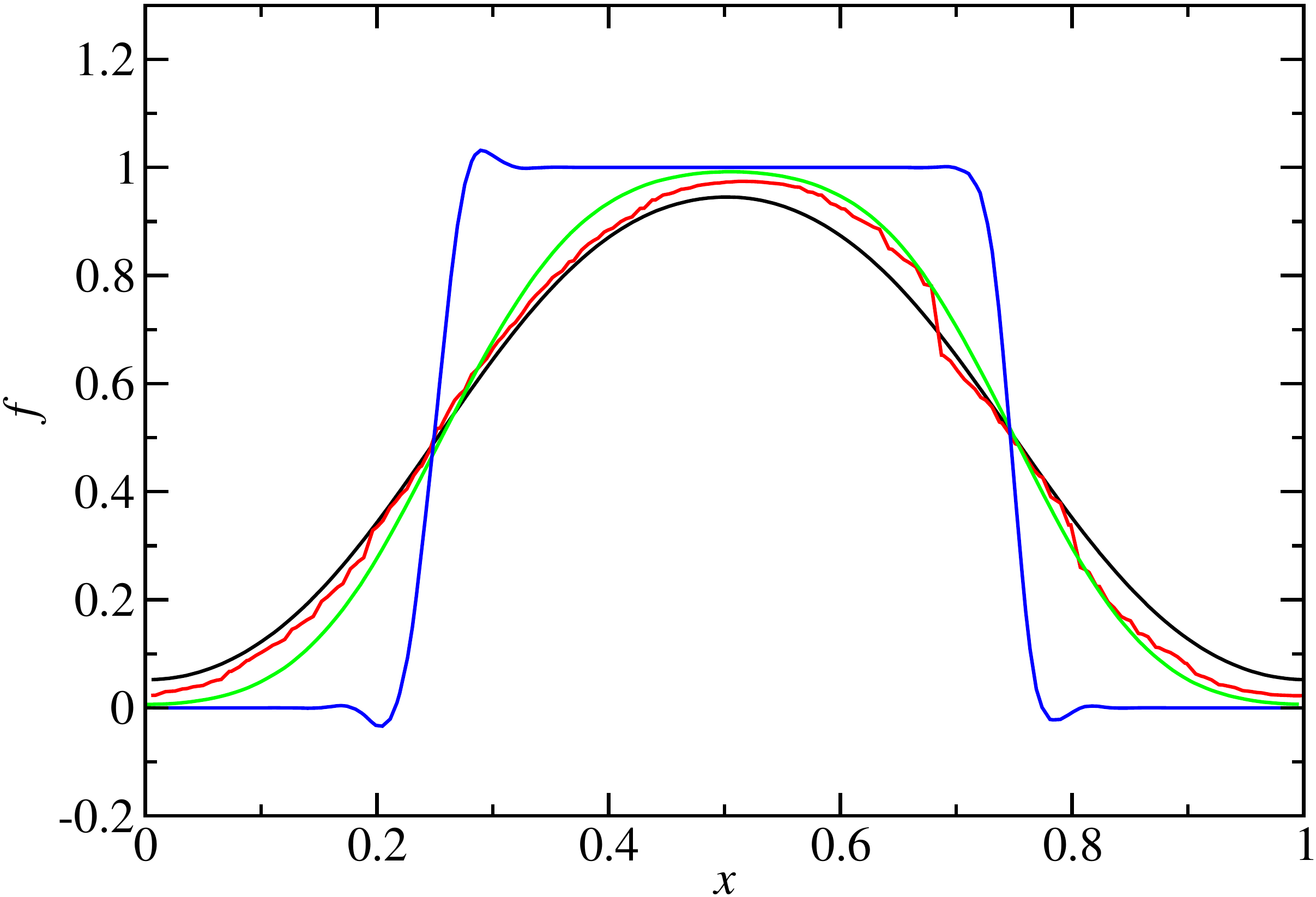}
  \end{minipage}
  \caption{Results for the moving top-hat function. Field after one
    traversal of the cell. Left: regular particle distribution, right:
    distorted. Black: $\delta$ projection, red: FLIP, green:
    mass-$\delta$, blue: full mass. \label{fig:1D_final}}
\end{figure}

In Figure \ref{fig:1D_final} we show the final profiles at time $T=1$,
at which the particles have traversed the system and come back to
their initial positions. At the left results are for a regular
particle set up, while at the right particles are disturbed
$\pm 40\% (\Delta x)$ about their initial positions. Notice that in
the former case the FLIP method is equivalent to the $\delta$ one.

The full mass method is seen to be the only one to preserve the
plateaus at $0$ and $1$, while the other methods spread out the
function that was initially sharp. This comes at the cost of
undershoots below $0$, and overshoots above $1$, close to the
interface, resembling Gibbs phenomenona.  This may be understood by
the fact that the assignment functions are not positive, as is obvious
from the requirement in Eq. (\ref{eq:ortho}). Indeed, for $j=i+1$ the
equation reads, for FEs in 1D:
\[
\int_{x_i}^{x_{i+1}} dx \phi_i(x) \psi_{i+1}(x) = 0,
\]
but since $\psi_{i+1}(x)$ is positive, it follows that $\phi_i(x)$ is
not.

In table \ref{table:1D_final} we provide measures of the accuracy of
the final profiles. First of all, we evaluate the relative change in
the integral of $A$:
\[
E_1:= \sqrt{
  \frac{
    \sum_\mu \left(
      (v_\mu A_\mu)(T=1) -
      (v_\mu A_\mu)(T=0) 
    \right)
  }{
    \sum_\mu (v_\mu A_\mu)(T=0) 
  } 
}
\]
This quantity should be null for methods that comply with
(\ref{eq:cons}).  Indeed we see this is satisfied to machine precision
by all methods for a regular particle arrangement. However, small
differences occur, as expected, when particles are distorted, except
for the FLIP method. Recall that the full mass method, while in
principle compliant with (\ref{eq:cons}), is implemented with an
approximate quadrature which will result in departures from this
property, see Appendix \ref{sec:quadrature}.

We also check the energy-like second moment of the profile, which
according to (\ref{eq:stab}) should always be a number that decreases
with each iteration, although of course, a slower decrease is
desirable. We have checked it indeed does, and measure its decrease by
its relative value at the last time step.
\[
E_2:=
\sqrt{ 
  \frac{
    \sum_\mu \left(
      (v_\mu A_\mu^2)(T=1) -  
      (v_\mu A_\mu^2)(T=1)
    \right)
 }{
    \sum_\mu (v_\mu A_\mu^2)(T=1) } 
}
\]
The full mass procedure is seen to produce a less distorted final
profile, as evident also in Figure \ref{fig:1D_final}.

Finally, we measure the relative $L_2$ distance between the final
profile and the initial one:
\[
L_2 :=
\sqrt{\frac{ 
    \sum_\mu \left(
      (v_\mu A_\mu)(T=1) - (v_\mu A_\mu)(T=0) 
    \right)^2 }{
    (v_\mu A_\mu)(T=0)
  }
},
\]
again confirming that the full mass projection is more accurate.

\begin{table}
  \centering
  \begin{minipage}{0.45\textwidth}
  \begin{tabular}{cccc}
    method & $E_1$ & $E_2$ & $L_2$
    \\
    \hline
    \hline
    $\delta$ &  $0$ &  $-29\%$ & $35\%$
    \\
    FLIP     &  $0$ &  $-25\%$ & $35\%$
           \\
    mass-$\delta$     &  $0$ &  $-21\%$ & $29\%$
    \\
    full mass     &  $0$ &  $0.85\%$ & $10\%$
    \\
    \hline
  \end{tabular}
  \end{minipage}
  \begin{minipage}{0.45\textwidth}
  \begin{tabular}{cccc}
    method & $E_1$ & $E_2$ & $L_2$
    \\
    \hline
    \hline
    $\delta$ &  $-0.24\%$ &  $-29\%$ & $35\%$
    \\
    FLIP     &  $0$ &  $-25\%$ & $35\%$
           \\
    mass-$\delta$     &  $0.7 \%$ &  $-20\%$ & $29\%$
    \\
    full mass     &  $-0.7\%$ &  $-4\%$ & $12\%$
    \\
    \hline
  \end{tabular}
  \end{minipage}
  \caption{Results for 1D spacing. Left table: regular, right: distorted\label{table:1D_final}}
\end{table}

\subsection{Zalesak's disk}
\label{sec:zalesak}

Let us consider a region with a colour field $\alpha$ that has a value
of $1$ for points inside a domain and $0$ for points outside and which
is simply advected:
\begin{equation}
  \frac{d A}{d t} = 0 \qquad   \frac{d \bfr}{d t} = u .
\end{equation}

The domain is a circle with a slot. The circle's radius is given a
value of $R=0.5$, while the slot was a width of $1/6$, and a height of
$5/6$. The simulation box is a $(-1.5,1.5)\times(-1.5,1.5)$ square, and the
number of nodes is set to $90 \times 90$, so that the mesh spacing is
$H=3/90=1/30$, the same value as in \cite{Idelsohn_2015}.  The time
step is $\Delta t=0.01$, which corresponds to
$\mathrm{Co}_H := u (\Delta t) /H \approx 0.94$ for nodes on the rim
of the disk.

The velocity field is a pure rotation:
\begin{align}
  u_x &= -\omega y\\
  u_y &=  \omega x ,
\end{align}
where $\omega=2\pi/\tau$, and the period of rotation is set to
$\tau=1$.  Periodic boundary conditions are used in this simulation,
but this fact is not really important since the only region that is
actually moved is within a circle of radius $1.4$, within a simulation
box of size $3 \times 3$.

For the mesh, and also for the particles in the mass methods, we use
linear FE functions, as in the previous 1D example. These are
piece-wise affine functions: pyramids of height $1$ at the node,
decreasing to $0$ at the neighbouring nodes. We employ the Delaunay
triangulation in order to determine these functions. For all mass
methods, a moving Delaunay triangulation must be maintained for the
projection from the mesh to the particles. The calculation of the
corresponding mass matrix, and its inversion, is only needed for the
full mass method.

\begin{figure}
  \centering
  \begin{minipage}{0.45\textwidth}
      \includegraphics[width=\textwidth]{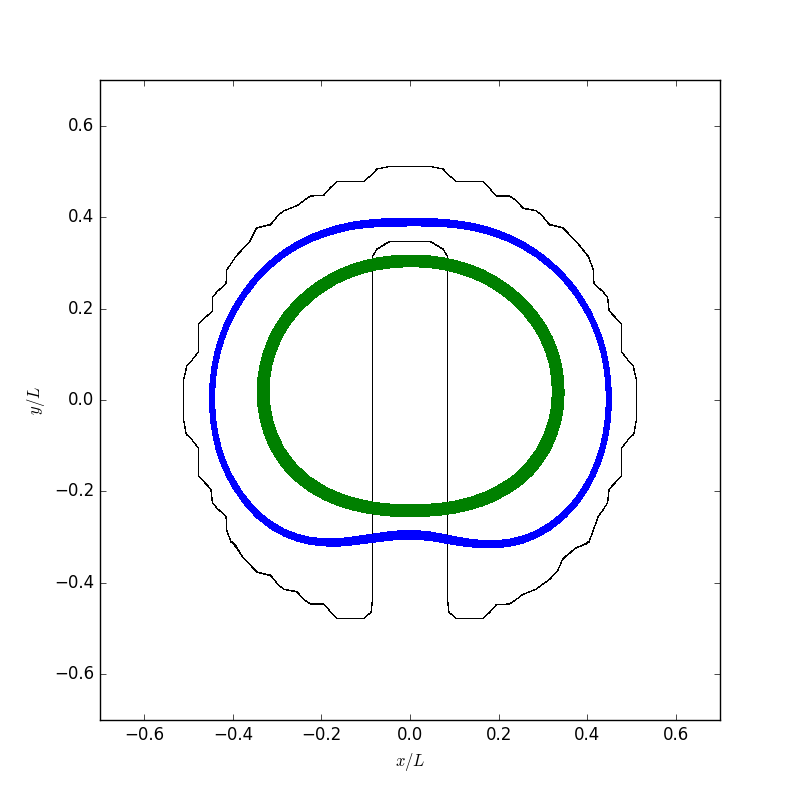}
  \end{minipage}
  \quad
  \begin{minipage}{0.45\textwidth}
      \includegraphics[width=\textwidth]{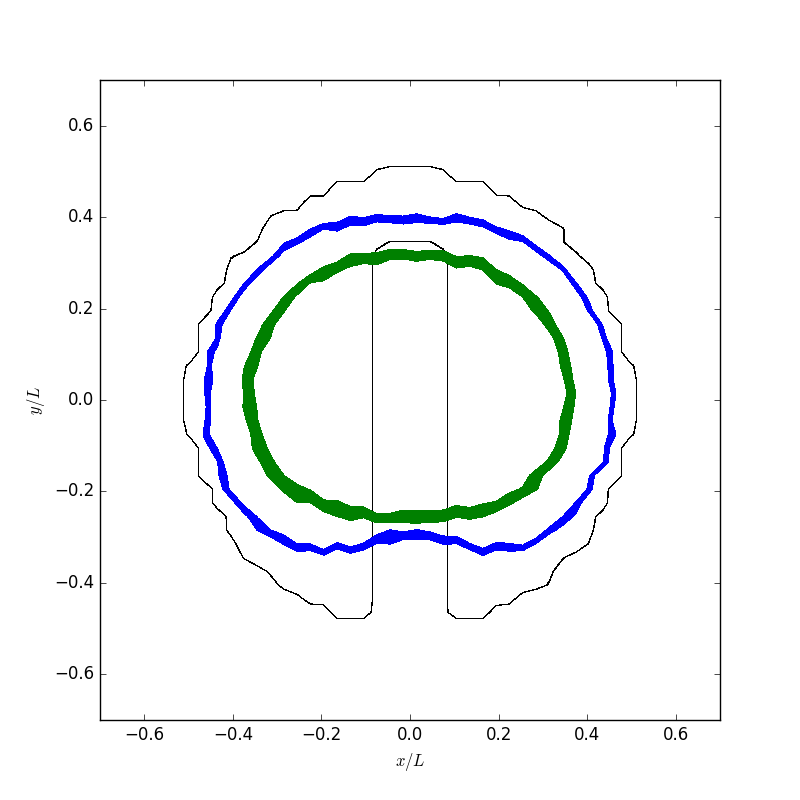}
  \end{minipage} \\
  \begin{minipage}{0.45\textwidth}
      \includegraphics[width=\textwidth]{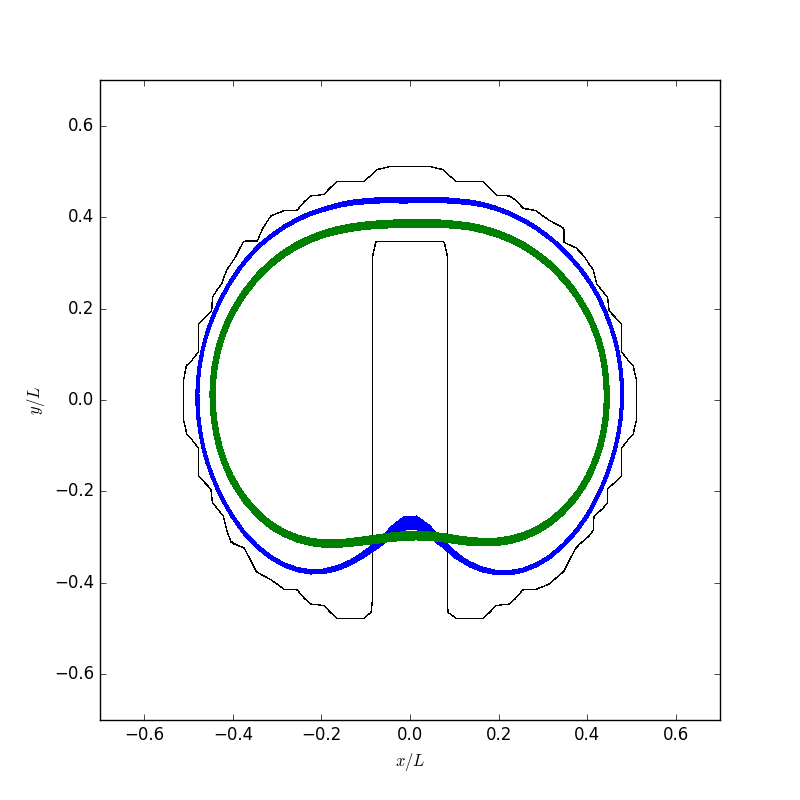}
  \end{minipage}
  \quad
  \begin{minipage}{0.45\textwidth}
      \includegraphics[width=\textwidth]{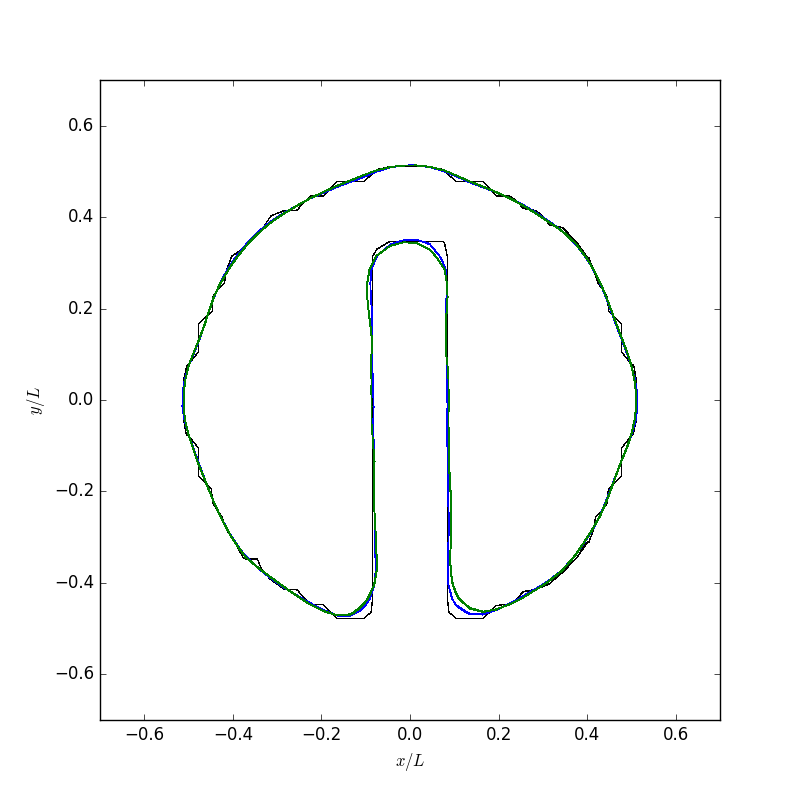}
  \end{minipage}
  \caption{Results for the rotation of Zalesak's disk.  Isocontours for
  $\alpha \in ( 0.49 , 0.51 )$. Initial field in black, after one
  rotation in blue, after two in green. Top left: $\delta$ method. Top
  right: FLIP. Bottom left: mass-$\delta$. Bottom right: full mass.
  \label{fig:zalesak}}
\end{figure}

Again, it would make little sense to project from and onto the mesh at
every time-step, but for benchmarking purposes. Results are given in
Figure \ref{fig:zalesak}, showing contour plots for values between
$0.49$ and $0.51$ of the $\alpha$ field on the mesh nodes.  We include
the initial contour, the contour after one revolution, $T=\tau$, and
after two revolutions, $T=2\tau$.

On the top left contours are shown for the $\delta$ procedure, and at
the top right, for the FLIP procedure. Both are seen to greatly smear
out the initial slab. At the bottom left, the full-$\delta$ procedure
is show. This time, some remains of the slab are seen after one
revolution. Finally, the full mass projection produces quite good
profiles even after two revolutions.



\begin{figure}
  \centering
  \includegraphics[width=0.5\textwidth]{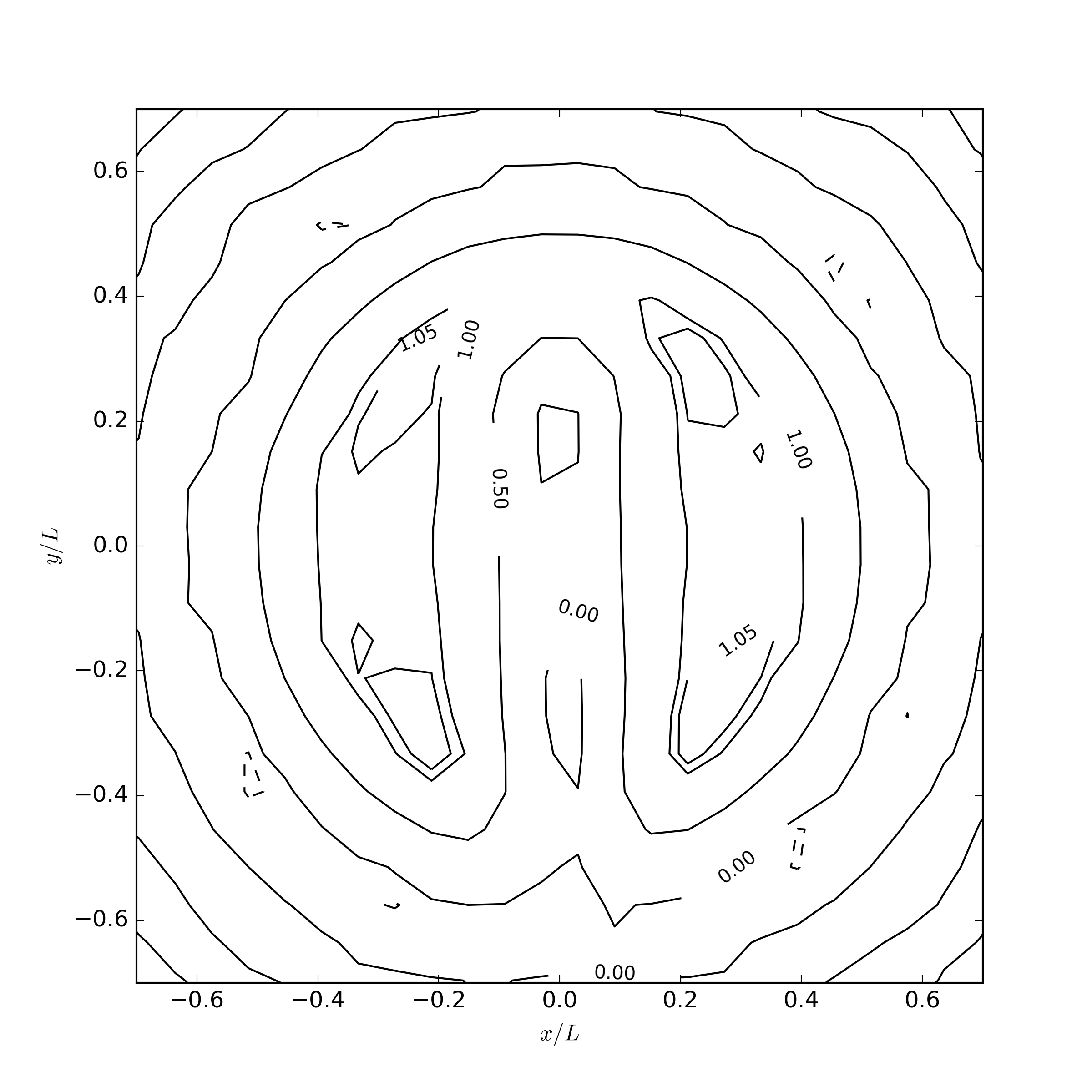}
  \caption{Results for the rotation of Zalesak's disk using the full
    mass method.  Isocontours for
    $\alpha \in ( -0.05, 0 , 0.5 , 1 , 1.05 )$ after two
    rotations. \label{fig:zalesak2}}
\end{figure}

As in 1D, the good results of the full mass method maintaining the
plateaus are accompanied by undershoots below $0$ and overshoots above
$1$. In Figure \ref{fig:zalesak2} we show more detailed contours for
the full mass procedure.  The isocontour for $\alpha=0.5$ is shown
again, but the $\alpha=0$ contour reveals a corona of slightly
negative values, as low as $-0.05$ approximately.  Values of $\alpha$
about $1.05$ are also seen in the inner regions.

We again employ the same error measures for our profiles. For the
relative $L_2$ distance we introduce a refinement, by comparing the
final profile and the one for a simulation in which the particle field
is simply advected. This way we subtract out the errors due to time
integration, by which particles' trajectories are not exactly
circles. These errors are quite small anyway.

\begin{table}
  \centering
  \begin{tabular}{cccc}
    method & $E_1$ & $E_2$ & $L_2$
    \\
    \hline
    \hline
    $\delta$ &  $-0.5\%$ &  $-29\%$ & $67\%$
    \\
    FLIP     &  $0.14\%$ &  $-61\%$ & $67\%$
    \\
    mass-$\delta$     &  $-0.5$ &  $-50\%$ & $61\%$
    \\
    full mass     &  $0.5\%$ &  $-9\%$ & $30\%$
    \\
    \hline
  \end{tabular}
  \caption{Results for the rotation of Zalesak's disk.\label{table:Zalesak}}
\end{table}

\subsection{Taylor-Green vortices}
\label{sec:TG}

The Taylor-Green vortex sheet is an analytic solution to the
Navier-Stokes equations for an incompressible Newtonian fluid:
\begin{equation}
  \frac{d \mathbf{u}}{d t} =  - \nabla p +  \nu \nabla^2 \mathbf{u}.
\end{equation}

The solution, with a periodic length of $L$, is the
velocity field
\begin{align}
\label{eq:TG_vel}
  \mathbf{u}_x &=  f(t) \sin (k x) \cos (k y) \\
  \mathbf{u}_y &= -f(t) \cos (k x) \sin (k y) ,\\
\end{align}
where $k=2\pi/L$, and the time-dependent prefactor function is given by
\[
f(t)=u_0 \exp\left( - 8 \pi^2 t^* /\mathrm{Re} \right) .
\]
The Reynolds number is defined as Re$:=u_0 L / \nu$, and the
dimensionless time is $t^* := t u_0/L$. We set $ u_0=1$, $L=1$, and
$\nu=0.005$, thus setting a Reynolds number of Re=$200$.

For the numerical solution of the Navier-Stokes equation, a standard
splitting approach is used \cite{codina_2001}.  The procedure is a
simple mid-point time integration, as in Ref. \cite{duque_2016b}. All
space derivatives are calculated on the mesh. As explained there, and
in Ref. \cite{Idelsohn_2015}, even this integrator, with $\Delta t^2$
accuracy, will result in a $\Delta t$ overall accuracy. This is
because the projection procedure causes a $\Delta t$ bottleneck in
the calculation. A possible remedy, proposed in
Ref. \cite{duque_2016b}, is to include higher order basis functions.
This idea is completely compatible with the ones put forward here, but
for the sake of simplicity we will not discuss them.



In order to quantify the accuracy of the different methods, 
the relative $L_2$ distance between the velocity field obtained by
simulation and its exact value is computed:
\begin{equation}
\label{eq:L2_error}
L_2:=
\sqrt{%
  \frac{
    \sum_{i=1}^N 
    V_i
    \left|
      \bfu(\bfr_i) - \bfu_i
    \right|^2
  }{
    \sum_{i=1}^N 
    V_i
    |\bfu(\bfr_i)|^2
  }
} ,
\end{equation}
where $\bfu(\bfr_i)$ is the exact velocity field as in
(\ref{eq:TG_vel}), evaluated on particle $i$ position. The same
measure may be evaluated for mesh nodes, with very similar
results.

\begin{figure}
  \centering 
  \includegraphics[width=0.7\textwidth]{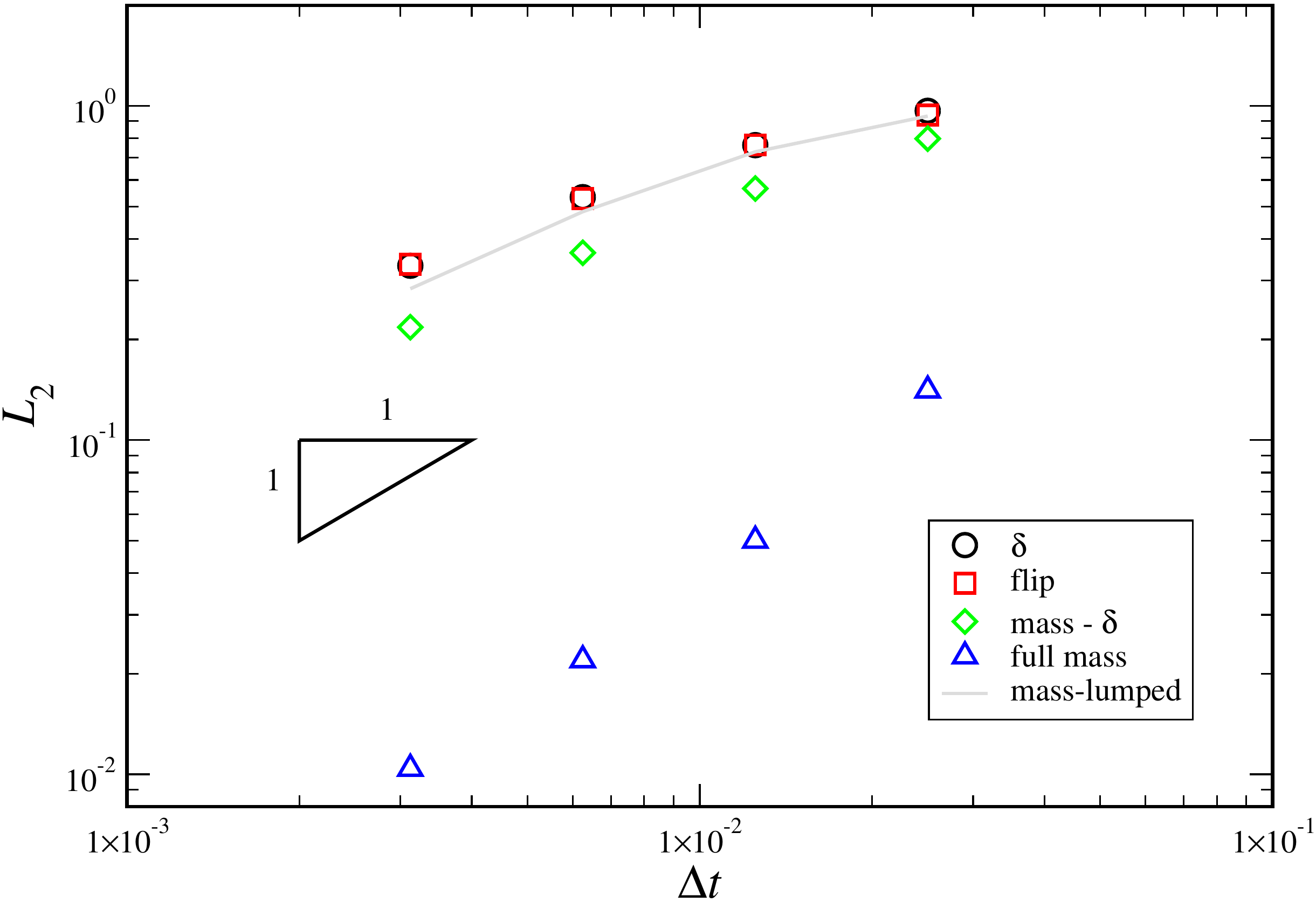}
  \caption{%
    $L_2$ error of the velocity  field at
    time $T^*=1$, versus time step $\Delta t$.  Black circles: $\delta$
    method, red squares: FLIP, green diamonds: mass-$\delta$, blue triangles:
    full mass. The triangle shows a $\Delta t^1$
    power-law.
    \label{fig:L2_x}
  }
\end{figure}

This error is expected to start at a very low value and increase
approximately linearly as the simulation proceeds. In order to compare
between methods, in Fig. \ref{fig:L2_x} the value of this error at
$T^*=1$ is plotted.  At this time, $f(T)=\exp(-8\pi^2 / 200 )=0.67 $,
so that the velocity field should have decreased to about $67\%$ of
its initial value.

The error for the velocity field (left subfigure) is seen to decrease
with $\Delta t$. The interparticle and mesh spacing is decreased as
$\Delta t$ does, in order to fix a Courant number of Co$_H=0.5$ (the
number of nodes and particles therefore increases quadratically).
Like in Zalesak's disk test, there are as many particles as nodes. The
particles are created at the beginning of the simulation and moved
according to the velocity field. As expected, the order of convergence
of the error agrees with a $\Delta t^1$ power in all cases. However,
results from the full mass procedure are much more accurate.


\begin{figure}
  \centering
  \includegraphics[width=0.7\textwidth]{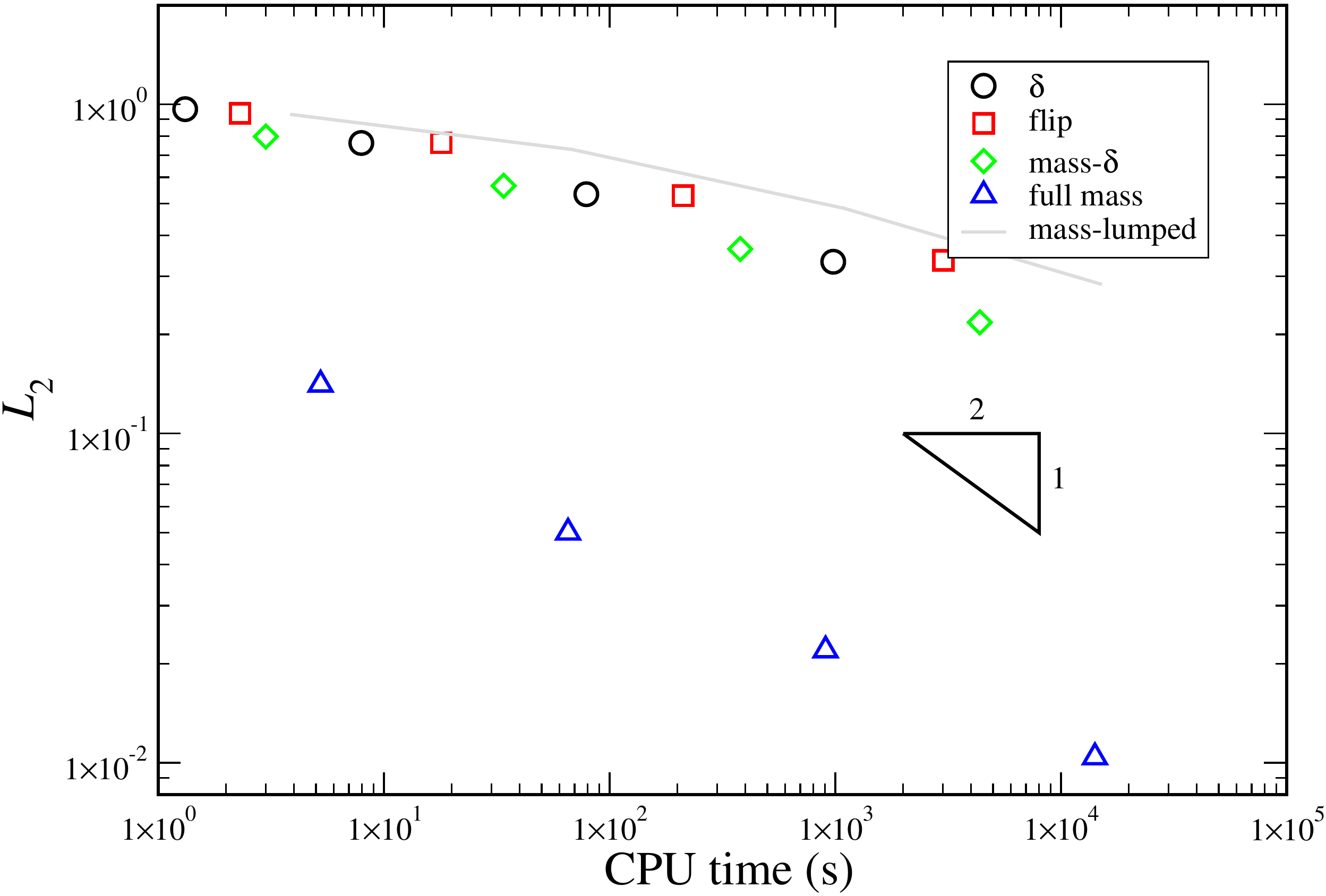}
  \caption{%
    $L_2$ error of the velocity  field at
    time $T^*=1$, versus CPU time in seconds  Black circles: $\delta$
    method, red squares: FLIP, green diamonds: mass-$\delta$, blue triangles:
    full mass. The triangle shows a $\Delta t^1$
    power-law.
    \label{fig:L2_x_vs_T}  }
\end{figure}

Despite its superior performance, the full mass is clearly more
costly computationally than the alternatives. It therefore seems
interesting to plot the error as a function of CPU time.
A bad scaling as the number of nodes and particles is increased
would make this procedure less appealing. However,
Fig. \ref{fig:L2_x_vs_T} makes clear that the full mass
procedure scales similarly to other procedures, with $L_2$ roughly
proportional to $\Delta t^{1/2}$.

The CPU run times are clearly depend on the machine used, but
a faster one, would likely make all simulations faster by a similar
factor. This would result in a horizontal translation of all the curves in a
logarithmic scale. Results do depend on the particular linear algebra
algorithm used, details can be found in Appendix \ref{sec:numerical}.

One may therefore conclude from these observations that the higher
computational cost of a full mass method will be compensated by
its higher accuracy.

\section{Conclusions}
\label{sec:conclusions}

We have described several assignment, or projection, procedures by
which field information may be transferred between particles and mesh.
We have focused on four procedures: the simplest $\delta$ method, the
FLIP method, the mass-$\delta$ method and the full mass method.

Each method is tested against several requirements: conservativity,
which makes sure that the total integral of a field is not changed;
stability, that ensures that the integral of the square of a field
decreases upon projection; and preservation of information, which
means that an assignment procedure, followed by interpolation, leaves
the field values invariant.

We have tested the method in 1D and 2D advection problems.
Conservativity is satisfied exactly by construction in the FLIP method,
and rather well satisfied in the other cases.  Stability is satisfied
in all four methods, but the full mass method is seen to be superior
in producing a smaller decrease. On the other hand, it also leads to
higher over- and undershoots at sharp interfaces.

Finally, the full mass method is clearly superior in our CFD simulations
of the Taylor-Green vortex sheet, where an approximate $10$-fold
increase in accuracy is achieved for the same simulation clock time.

The final conclusion is that the full mass method should be seriously
considered as an assignment procedure, despite its inherent higher
computational complexity.

\section*{Acknowledgements}

We wish to thank Prof. David Le Touz\'e for his suggestions regarding
the FLIP method. We also thank Prof. Michael Schick for hosting a
research stay, during which this work was finished. The research
leading to these results has received funding from the Ministerio de
Econom\'{\i}a y Competitividad of Spain (MINECO) under grants
TRA2013-41096-P ``Optimizaci\'on del transporte de gas licuado en
buques LNG mediante estudios sobre interacci\'on fluido-estructura''
and FIS2013-47350-C5-3-R ``Modelizaci\'on de la Materia Blanda en
M\'ultiples escalas''.

\bibliographystyle{plain}
\bibliography{PiC,sme_pdes,comp_geom}

\appendix

\section{Numerical methods}
\label{sec:numerical}

For all computational geometry procedures the CGAL 4.7 libraries
\cite{CGAL} are used.  In particular, the 2D Periodic Delaunay
Triangulation package, overloading the vertex base to contain the
relevant fields, and the face base to contain information relevant to
the edges.

The Eigen 3.0 linear algebra libraries \cite{Eigen} are also employed.
For the small linear algebra problem involved in the calculation of
the $A$ coefficients, SVD is used, with automatic rank detection. For
the large problems involved in the Galerkin procedure, the sparse
matrix package is used. The linear systems are solved iteratively for
pFEM, by the BiCGSTAB method.  For projFEM a direct method is
employed, with best results obtained using the CHOLMOD \cite{cholmod}
routines of the suitesparse project (through Eigen wrappers for
convenience, class CholmodSupernodalLLT). Slightly worse results are
obtained with eigen's build-in SimplicialLDLT class.

Our computations took place on a 4-core Pentium 4 machine with 16 Gb
RAM.
The code employed, named polyFEM, may be found in \cite{polyFEM}
under an open source license.


\section{Quadrature}
\label{sec:quadrature}

As explained in the main text, the ``mass'' assignments involve
integrals as in Equation (\ref{eq:proj}).

\begin{figure}
  \centering
  \includegraphics[width=0.45\textwidth]{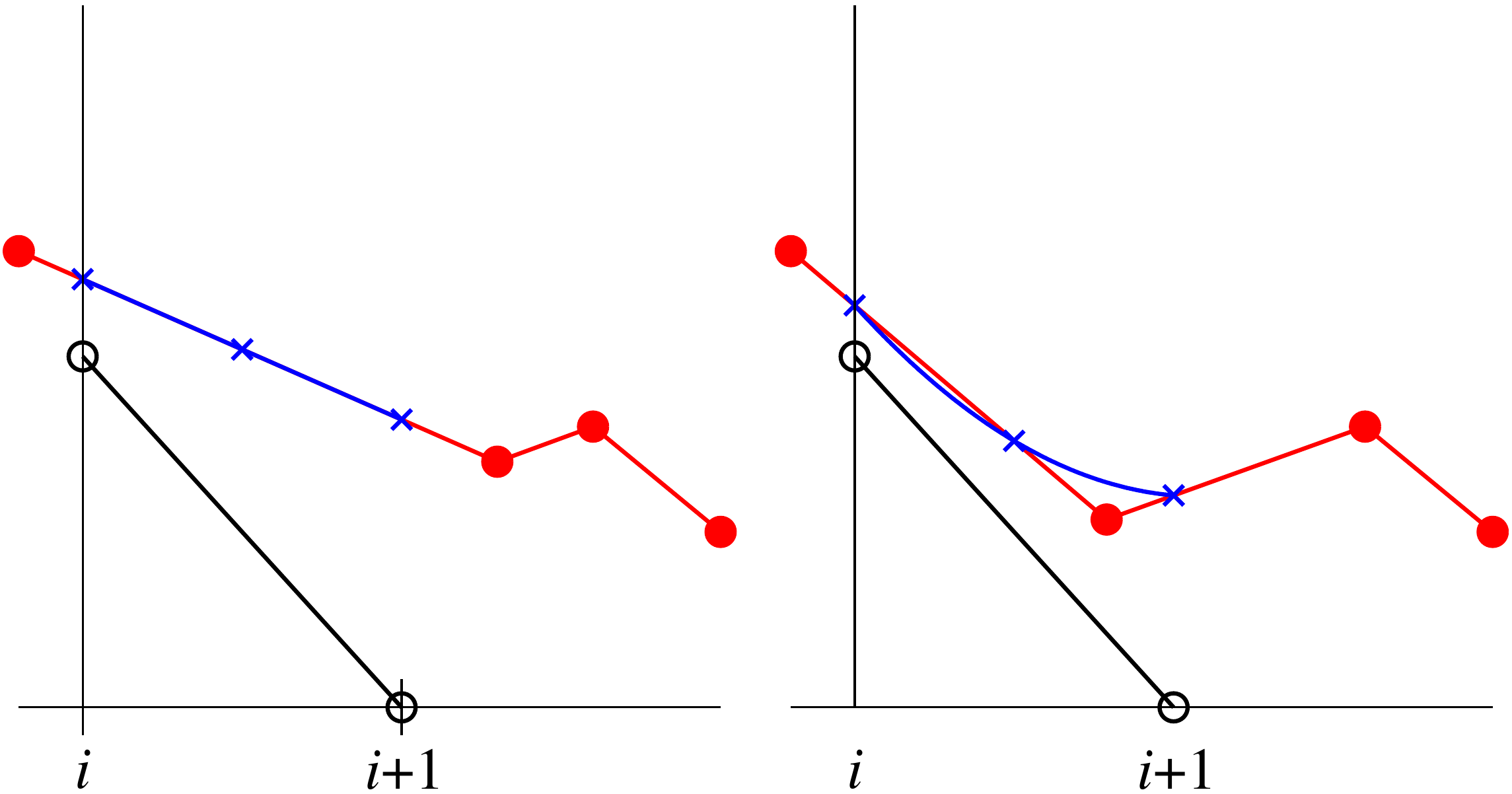} $\qquad$
  \includegraphics[width=0.45\textwidth]{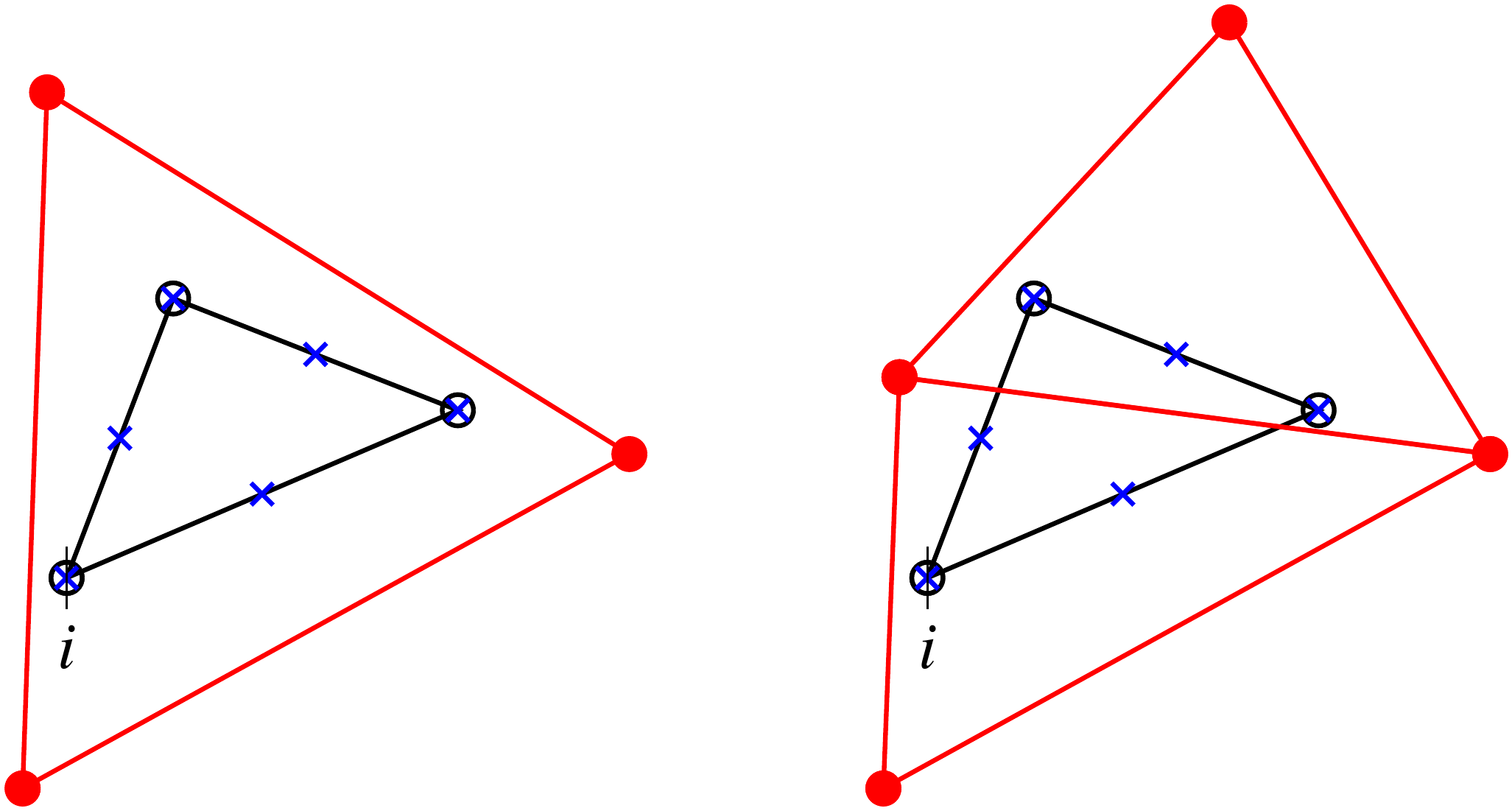}
  \caption{%
    Illustration of the quadrature used in 1D (left two figures), and
    2D (right two figures). Black empty circles: position of nodes. Red
    full circles: position of circles. Blue crosses: quadrature
    points.  In 1D, $\phi_i(x)$ is shown as a black line, and $A(x)$
    as a red line. The quadratic approximation to the latter is shown
    as a blue line.
    \label{fig:quadrature}}
\end{figure}

In this work, the $A(\bfr)$ is a piece-wise linear function, that
connects the values of at the particles $A_\mu$. The fact that the
locations of mesh nodes and particles do not match makes the integral
somewhat cumbersome to evaluate.  We have therefore implemented a
simple quadrature rule. It is best visualized in 1D, see Figure
\ref{fig:quadrature}. For the interval between node $i$ and $i+1$
function $A(x)$ is evaluated at $x_i$, $x_{i+1}$ and the position in
between. The resulting three values are then fit to a parabola, whose
overlap integral (functional scalar product) with $\phi_i$ is trivial
to evaluate. As shown in the Figure, if no particles lie on the
interval, this approximation is exact: the parabola will degenerate
to a line in this case. If some particle lies in the interval the
result will be, in general, an approximation.  For each node $i$ there
will be an additional integration from $i-1$ to $i$.

The same procedure is carried out in 2D, as also depicted in Figure
\ref{fig:quadrature}. The view is from above, and for simplicity only
2D points are drawn, not functions. Function $A(\bfr)$ is evaluated at
the nodes of the triangle that connects node $i$ and two of its
neighbours in the triangulation. In addition, it is also evaluated at
the mid-points of each segment. With these six values, an
interpolating quadratic function is found, whose overlap integral with
$\phi_i$ is trivial. Again, if no particles happen to lie in the
triangle the procedure is exact, but it is approximate if one or more
particles do lie in the triangle. For each node $i$ integration must
be performed over all its other incident triangles.

An equivalent procedure is applied in the inverse procedure of
\ref{eq:proj2}.

\end{document}